# Energy-Efficient Relay Selection and Optimal Relay Location in Cooperative Cellular Networks with Asymmetric Traffic

YANG Wei[1,2], LI Li-hua[1,2], SUN Wan-lu[1,2]

1. Key Lab. of Universal Wireless Commun., Beijing University of Posts and Telecom. (BUPT), Ministry of Education, Beijing, 100876, China
2. Wireless Technology Innovation Institute (WTI), BUPT, Beijing, 100876, China

**Abstract**

Energy-efficient communication is an important requirement for mobile relay networks due to the limited battery power of user terminals. This paper considers energy-efficient relaying schemes through selection of mobile relays in cooperative cellular systems with asymmetric traffic. The total energy consumption per information bit of the battery-powered terminals, i.e., the mobile station (MS) and the relay, is derived in theory. In the Joint Uplink and Downlink Relay Selection (JUDRS) scheme we proposed, the relay which minimizes the total energy consumption is selected. Additionally, the energy-efficient cooperation regions are investigated, and the optimal relay location is found for cooperative cellular systems with asymmetric traffic. The results reveal that the MS-relay and the relay-base station (BS) channels have different influence over relay selection decisions for optimal energy-efficiency. Information theoretic analysis of the diversity-multiplexing tradeoff (DMT) demonstrates that the proposed scheme achieves full spatial diversity in the quantity of cooperating terminals in this network. Finally, numerical results further confirm a significant energy efficiency gain of the proposed algorithm comparing to the previous best worse channel selection and best harmonic mean selection algorithms.

## 1   Introduction

Cooperative relaying is a promising technology that can improve the performance of a wireless system via a number of mechanisms, such as increased spatial diversity or beamforming effects [1]-[3]. There is no doubt that these mechanisms can increase the spectral efficiency and the total throughput of the whole system. However, using such techniques typically implies accepting higher power consumption not only of the transceiver but also of the complete radio access network. Therefore, relay techniques that improve the energy-efficiency are essential to the cellular networks, as they are not only good for the environment but also make commercial sense for operators and support sustainable, profitable businesses.

Since the relays consume the system resources and power, the total energy-efficiency of the relaying techniques may be limited. Therefore, one interesting issue in the relaying system is to determine whether a two-hop transmission is necessary. And it is also important to select a relay among available candidates to maximize cooperation benefits for the user or for the whole system [4]. Relay selection is widely studied in previous works. Nosratinia et al [5] demonstrate that relay selection techniques can capture maximum diversity in the number of cooperating nodes, while each node only knows its own receive channel state. Madan et al [6] consider selecting relays by minimizing the total power consumption. Tannious et al [3] propose an ITRS protocol, which employs hybrid-ARQ with packet combining at the destination and includes a limited-feedback handshake for relay selection, achieving the multiple-input single-output (MISO) diversity-multiplexing tradeoff (DMT) bound. Bletsas et al. [7] introduce a decentralized, opportunistic relaying scheme to select the best relay based on instantaneous end-to-end channel conditions, which also achieves the MISO DMT bounds.

All of the above-mentioned works focus on single direction transmission, i.e., from the source to the destination. However, in most cellular networks, the uplink and downlink may be in deep fading at the same time. In this scenario, relays are needed for both the uplink and the downlink, and the same relay can be adopted if the channel reciprocity exists. The base station (BS) and the mobile station (MS) can act interchangeably as the source and the destination, with the selected relay offering help for both two communications (see Fig. 1). Meanwhile, the traffic loads of the uplink and the downlink have been changing to be asymmetric to support new multimedia communication services. For example, the downlink traffic load may be larger than that in the uplink when the MS downloads some files from the BS. In this scenario, since the relay transmits signal to MS more often than to BS, the MS-relay channel may have more influence than the BS-relay channel over relay selection. Motivated by this, the traffic load condition of the uplink and downlink is taken into account during the selection of relay.

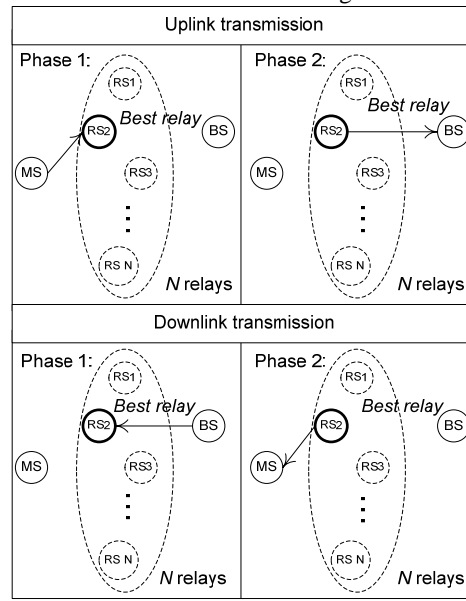

**Fig. 1** We consider communication between BS and MS via one mobile relay. In the uplink transmission, the relay forwards the signal from MS to BS, and in downlink, the same relay forwards the signal from BS to MS.

In this paper, an energy-efficient relaying scheme is proposed through selection of mobile relays. The proposed scheme has several improvements compared to the conventional schemes.

- Relay selection is performed for the uplink and downlink jointly, and the selection overhead can be reduced significantly.
- Traffic load conditions of the uplink and downlink are considered during the selection. Thus the scheme can be more convenient to adapt the asymmetric traffic services in the next-generation wireless communications.
- The relay which minimizes the total energy consumption per information bit of battery terminals is selected. Simulation results show that the energy-efficiency of the proposed scheme is improved over previous best worst channel and best harmonic mean selection schemes.

The relay selection results are illustrated in terms of the network geometry for a range of potential relay locations, afterwards. Energy-efficient cooperation regions are defined and studied, and the optimal relay location is found for cooperative cellular systems with asymmetric traffic. Our results reveal that the MS-relay and the relay-BS channels have different influence over relay selection decisions for optimal energy-efficiency.

The remainder of this paper is organized as follows. In Section 2, we describe the system model. The proposed relay selection scheme is presented in Section 3. In Section 4, the optimal relay location in cooperative cellular systems is studied. The performance evaluations are presented in Section 5. Finally, Section 6 concludes the work of the paper.

## 2　System model

We consider a half-duplex dual-hop communication scenario in a fading environment with one MS, one BS, and a set $N$ decode-and-forward (DF) relays, as depicted in Fig. 1. During the first hop, the source transmits its information to the relays and destination, while in the second phase, one relay (assume that each time, at most one relay is used) decodes and forwards the received signal to the destination. The destination receiver combines the messages from the source and the relay using optimal diversity combining. In cooperative cellular networks, MS and BS act interchangeably as the source and destination in the uplink and downlink transmission, with relay offering help for both the uplink and downlink communications.

### 2.1　Channel model

The power gains of the MS-relay and relay-BS channels are denoted as $h_{1,j}$ and $h_{2,j}$, $j=1,2,\cdots N$ respectively. The channel power gain $h_{i,j}$ captures the effect of large scale fading and Rayleigh fading, which can be expressed as

$$h_{i,j} = K \cdot d_{i,j}^{-p} \cdot \upsilon_{i,j}, \quad i \in \{1,2\}, \quad j=1,2,\cdots N \tag{1}$$

where $K$ is a constant defined by the antenna gain, carrier frequency and other system parameters, $p$ is the pathloss component, $\upsilon_{i,j}$ is an exponentially distributed random variable, $d_{1,j}$ is the distance between the MS and relay $j$, and $d_{2,j}$ is the distance between relay $j$ and the BS. The channel power gain and distance between MS and BS are denoted as $h_D$ and $D$, respectively. The uplink and downlink of each channel are assumed reciprocal. At all nodes, the additive white Gaussian noise (AWGN) has a power spectral density of $N_0$. All transmissions in the system have a bandwidth of $B$ Hz.

One important parameter of the system is the spectral efficiency [2]. We assume the transmission schemes are parameterized by the MS-BS spectral efficiency $R$ bits/s/Hz. As the transmission through a relay needs two transmission phases, the spectral efficiency of each transmission phase should be no less than $2R$ bits/s/Hz to get an end-to-end $R$ bits/s/Hz spectral efficiency. For simplicity, the end-to-end spectral efficiency of the uplink and downlink are assumed identical, which are $R$ bits/s/Hz.

2.2 Asymmetric traffic model

To model the asymmetric traffic condition in the system, the traffic asymmetry factor $\zeta$ are introduced, which gives the ratio of the traffic load $L_{UL}$ in the uplink to the total traffic load $L_{total}$. $L_{total}$ is the sum of uplink traffic load $L_{UL}$ and the downlink traffic load $L_{DL}$. The traffic load $L_{total}$, $L_{UL}$ and $L_{DL}$ are defined as the number of generated bits in a communication.

2.3 Energy Consumption model

The power consumption of a user terminal in transmission mode comprises the transmission signal power $P_T$ and the circuit power consumption $P_C$ in the whole signal path [8]. The transmission signal power $P_T$ depends on the data transmission rate $R$. Thus, the overall power used for data transmission is given as $P = P_T + P_C \leq P_{max}$, where $P_{max}$ is the maximum available battery power at the user terminals.

In the scope of energy-efficiency estimation methodologies, a measure usually taken for the comparison of radio transmission technologies is the energy consumption per information bit [9]. However, in cooperative cellular networks, a comparison of the sheer energy consumption is not suitable, because certainly the MS, relay and BS have different power consumption challenges. In this paper, we only account for the energy consumed by MS and the relay, and ignore the energy consumption of the BS. This is reasonable as the MS and relay are all powered by batteries while the BS is always powered by a fixed line. Thus, the total energy consumption of battery powered terminals per information bit in cooperative communication can be expressed as

$$E_{coop} = \left(P_M^U T_M^U + P_R^U T_R^U + P_R^D T_R^D\right)/L_{total} = \left(\zeta\left(P_M^U + P_R^U\right) + (1-\zeta)P_R^D\right)/2RB \tag{2}$$

where $P_M^U$ and $P_R^U$ is the power of the MS and the relay used for uplink data transmission, $P_R^D$ is the power of the relay used in the downlink, and $T_M^U$, $T_R^U$ and $T_R^D$ are the corresponding data transmission time.

## 3 Joint Uplink and Downlink Relay Selection

This section considers the energy-efficient relay selection scheme. The trellis coded M-ary Quadrate Amplitude Modulation (MQAMs) are used as the MCS.

3.1 Selection criterion

The bit error rate (BER) for coherently detected MQAM with Gray mapping over Rayleigh fading channel is approximated by [10]

$$P_e(\gamma) \approx 0.2 \exp\left(-\frac{1.5\gamma}{M-1}\right) \tag{3}$$

where $\gamma$ is the signal-to-noise ratio (SNR) at the receiver. The SNR is defined as $\gamma = (P_T h)/(N_0 B)$, where $h$ is the channel power gain. For a given BER $P_e$, the required SNR can be determined by Eq. (3). Consequently, the required transmit power will be

$$P_T = \frac{2\ln(5P_e)N_0 B}{3h}(1-M) \tag{4}$$

It is well known [11] that the forward error correction code (ECC) can reduce the required value of $E_b/N_0$ by a factor of the coding gain $G_C$ under the same performance requirement (specified as the BER), where $E_b$ refers to the received energy per information bit. Due to the embedded ECC, the required power for trellis coded MQAM $P_{TC}$ is given as

$$P_{TC} = \frac{P_T \eta_C}{G_C} = \frac{2\eta_C \ln(5P_e)N_0 B}{3h G_C}\left(1 - 2^{R/\eta_C}\right) \tag{5}$$

where $\eta_C$ is the code rate of the trellis encoder. For a two hop cooperative transmission using DF protocol, the end-to-end BER is given by

$$P_e = 1 - (1 - P_e(\gamma_1))(1 - P_e(\gamma_2 + \gamma_D)) \tag{6}$$

where $\gamma_1$, $\gamma_2$ and $\gamma_D$ are the SNR of the first hop, the second hop and the direct link, respectively, $P_e(\gamma_1) \triangleq P_{e,1}$ and $P_e(\gamma_2 + \gamma_D) \triangleq P_{e,2}$ are the BER of the first and the second hop. According to [12], the optimal power allocation scheme for DF relaying satisfies $\gamma_1 = \gamma_2 + \gamma_D$. Therefore, the required target BER of the first hop and the second hop are identical, i.e., $P_{e,1} = P_{e,2} = 1 - \sqrt{1-P_e}$.

Consequently, the expression of total energy consumption per information bit for the cooperative communication between the MS and the BS through relay $i$ is given by

$$E_{coop,i} = \frac{\zeta(P_{M,i}^U + P_{R,i}^U) + (1-\zeta)P_{R,i}^D}{2RB} = \frac{\zeta}{2RB}\left(\frac{F(2R)}{h_{1,i}} + \frac{F(2R)(1-h_D/h_{1,i})}{h_{2,i}} + 2P_C\right) + \frac{1-\zeta}{2RB}\left(\frac{F(2R)(1-h_D/h_{2,i})}{h_{1,i}} + P_C\right) = \frac{F(2R)}{2RB}\left(\frac{1}{h_{1,i}} + \frac{\zeta}{h_{2,i}} - \frac{h_D}{h_{1,i} \cdot h_{2,i}}\right) + \frac{(1+\zeta)P_C}{2RB} \tag{7}$$

where $F(R) = \frac{2\eta_C N_0 B}{3G_C}(2^{R/\eta_C} - 1)\ln\left(\frac{1+\sqrt{1-P_e}}{5P_e}\right)$. Finally, the relay selection method is described as

$$i^* = \arg\min_{i \in \{1,2,\cdots N\}}\left(\frac{1}{h_{1,i}} + \frac{\xi}{h_{2,i}} - \frac{h_D}{h_{1,i}h_{2,i}}\right) \tag{8}$$

subject to $\frac{F(2R)}{h_{1,i}} + P_C \leq P_{max}$ and $\frac{F(2R)(h_{1,i} - h_D)}{h_{1,i}h_{2,i}} + P_C \leq P_{max}$. We call this criterion as the *Minimum Energy Criterion*.

Note that MS can also communicate with BS directly without the help of any relay if the channel gain of the direct link is strong enough. The energy consumption per bit required for direct transmission between MS and BS for successful transmission at data rate $R$ is given by

$$E_{direct} = -\frac{2\zeta\eta_C N_0 \ln(5P_e)}{3h_D G_C R}(2^{R/\eta_C} - 1) + \frac{\zeta P_C}{RB} \tag{9}$$

In *Minimum Energy Criterion*, no relay is selected if $E_{direct} \leq E_{coop,i^*}$. In this case, BS communicates with MS directly.

### 3.2 Joint uplink and downlink relay selection Scheme

This subsection presents a relay selection protocol for cooperative cellular systems with asymmetric traffic, called Joint Uplink and Downlink Relay Selection (JUDRS). The detailed procedure of JUDRS is as follows.

**Step 1:** MS broadcasts a RTS1 packet to the relays and BS using a fixed transmission power $P_{max} - P_C$. Each relay hears the RTS1 packet and estimates the power gain of the channel $h_{1,i}$ between MS and itself. Depending on the channel states, only a subset $\Gamma$ of the $N$ relays can be chosen as candidate relays, defined by $\Gamma \triangleq \{i \in \{1,2,\cdots N\}: (P_{max} - P_C)h_{1,i} \geq F(2R) \triangleq g_{th1}\}$. BS also measures the channel gain of the direct link $h_D$.

**Step 2:** The relays in $\Gamma$ send RTS2 packets to the BS along with the channel quality indicator (CQI) using power $P_{max} - P_C$.

**Step 3:** BS estimates the channel gain $h_{2,i}$ between it and relay $i$, $i \in \Gamma$. The relay can be helpful to the transmission between MS and BS only if $(P_{max} - P_C)h_{2,i} \geq F(R)(1 - h_D/h_{1,i}) \triangleq g_{th2}$. These relays form the candidate relay set $\Sigma$.

**Step 4:** BS selects the best relay from $\Sigma$ under the *Minimum Energy Criterion*, and broadcasts the index of the best relay along with the channel power gain $h_D$ of the direct link. $h_D$ is used in the selected relay to calculate the transmission power in the uplink and downlink communication.

**Step 5:** MS and BS communicate with each other in uplink and downlink via the selected relay. The transmission power of MS and the selected relay can be calculated as

$$P_M^U = \frac{g_{th1}}{h_{1,i}}, \quad P_R^U = \frac{g_{th2}}{h_{2,i}}, \quad P_R^D = \frac{F(R)}{h_{1,i}}\left(1 - \frac{h_D}{h_{2,i}}\right) \tag{10}$$

If the direct link is selected, MS and BS communicate directly with each other.

Fig. 2 shows the flow chart of the JUDRS scheme. It can be seen that, the relay for the uplink and downlink transmission is selected jointly in on selection. Thus the relay selection overhead is significantly reduced compared with conventional schemes in which the relays are selected separately for uplink and downlink.

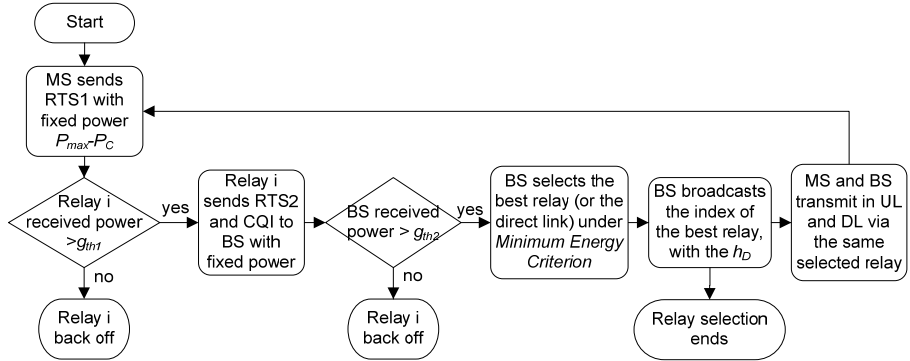

**Fig. 2** The flow chart of the JUDRS scheme

## 4 Optimal relay location and energy efficient cooperation region

To illustrate the results of energy consumption in terms of network geometry for a range of potential partner locations, we map the channel power gains to inter-node distances, as is used as in Eq. (1), and the Rayleigh fading is ignored. In this section, $d_{1,i}$ and $d_{2,i}$ are denoted by $d_1$ and $d_2$ for simplicity.

**Definition 1 (Cooperation Energy-Gain):** In cooperative cellular networks, cooperation energy-gain is defined as the percentage of energy saving achieved by cooperation to get the same spectral efficiency as the direct transmission between MS and BS,

$$E_{saving\_coop} = \frac{E_{direct} - E_{coop}}{E_{direct}} \times 100\% \tag{11}$$

where $E_{coop}$ and $E_{direct}$ are defined in Eq. (7) and Eq. (9) respectively. The energy saving $E_{saving\_coop}$ achieved by cooperation can be obtained as

$$E_{saving\_coop} = \frac{\frac{2\eta_C N_0 \ln(5P_e)}{3G_C RK}\left(2^{R/\eta_C}-1\right)\left(\ln\left(\sqrt{1+\sqrt{1-P_e}}\right)\left(2^{R/\eta_C}+1\right)\left(d_1^p + \zeta d_2^p - d_1^p d_2^p D^{-p}\right) - \zeta D^p\right) - \frac{(1-\zeta)P_C}{2RB}}{-\frac{2\eta_C N_0 \ln(5P_e)}{3G_C RK}\left(2^{R/\eta_C}-1\right)\zeta D^p + \frac{\zeta P_C}{RB}}$$

$$\overset{(a)}{\approx} \frac{H(R)\left(\zeta - \ln\left(\sqrt{2}\right)\left(2^{R/\eta_C}+1\right)\left(\tilde{d}_1^p + \zeta \tilde{d}_2^p - \tilde{d}_1^p \tilde{d}_2^p\right)\right) - \frac{(1-\zeta)P_C}{2RB}D^{-p}}{H(R)\zeta + \frac{\zeta P_C}{RB}D^{-p}} \tag{12}$$

where $\tilde{d}_1 = d_1/D$, $\tilde{d}_2 = d_2/D$, and $H(R) = -\frac{2\eta_C N_0 \ln(5P_e)}{3G_C RK}\left(2^{R/\eta_C} - 1\right)$. Here (a) holds because $1 + \sqrt{1-P_e} \approx 2$, if $P_e \ll 1$. It is evident that transmission energy is a function of inter-node distances. Hence, the cooperation energy-gain changes accordingly with the location of the relay.

**Definition 2 (Energy-efficient cooperation region):** For cooperative cellular networks, the energy-efficient cooperation region is defined as the locations such that cooperating with the relays in those locations requires less transmission energy than the direct communication between the MS and the BS. In other words, if a relay is inside the energy-efficient cooperation region, the energy saving achieved by cooperation $E_{\text{saving\_coop}}$ is greater than 0.

**Definition 3 (Optimal Relay location):** For cooperative cellular networks, the optimal relay location is defined as the location of relay that maximizes the energy saving, i.e., the optimal relay location is defined by

$$(\tilde{d}_1^{\text{opt}}, \tilde{d}_2^{\text{opt}}) = \arg\min_{\tilde{d}_1, \tilde{d}_2}\left(E_{\text{saving\_coop}}\right) \qquad (13)$$

Subject to $\tilde{d}_1 + \tilde{d}_2 \geq 1$.

Unless otherwise specified, the parameters used in the following contexts are shown in Table 1, where $f_c$ is the frequency of the central carrier.

Table 1　Parameters

| Parameter | Value |
| --- | --- |
| $P_{\max}$ | 33 dBm [13] |
| $K$ | −128.1dB [14] |
| $B$ | 180 KHz |
| $f_C$ | 2.0 GHz |
| $N_0$ | −171 dBm/Hz [9] |
| $p$ | 3.76 [14] |
| $G_C$ | 4.7 dB |
| $\eta_C$ | 2/3 |
| $P_e$ | $10^{-4}$ |
| $P_C$ | 20 dBm [13] |

Fig. 3 is a contour graph of energy saving per information bit achieved by cooperation. The BS is located 500m away from the MS, which is located at the origin. Fig. 3 confirms that the closer a potential relay is to both the MS and the BS, the higher the energy savings achieved from cooperation. And any relay located outside the outmost circle in Fig. 3 will not be helpful to the transmission between the MS and the BS in terms of energy-efficiency.

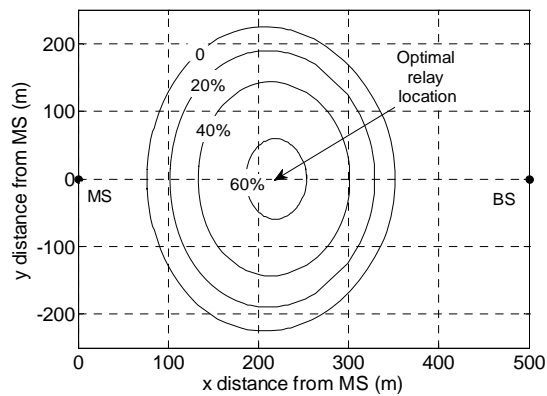

**Fig. 3**　Energy-efficient cooperation regions for a certain energy $E_{\text{saving\_coop}}$, $D = 500$ m, $\zeta = 0.5$

Interestingly, the energy-efficient cooperation region derived in this paper is much smaller than the cooperation region developed in [4], which is based on minimum transmission frame error rate (FER). In another word, in order to improve the energy-efficiency, the restriction upon the relay location is much stricter than that for spectral efficiency improvement. Moreover, the best relay location lies closer to the MS than to BS. By contrast, in [15], it is suggested that a relay located at equal distances from source and destination has a better performance. Therefore, relay selection for cooperative cellular system with asymmetric traffic is distinctly different from the relay selection for one-direction communication.

Fig. 4 shows the energy-efficient cooperation regions for different asymmetric traffic factor $\zeta$. It is illustrated that the size of energy-efficient cooperation region increases with the increasing of $\zeta$ (recall that $\zeta$ is the ratio of uplink traffic load to the total traffic load). This is reasonable as the energy-consumption of the MS in traditional non-cooperative transmission reduces proportionally with the decrease of the uplink traffic load. Meanwhile the energy consumption of the user terminals in cooperative transmission does not decease proportionally with $\zeta$, as the energy consumed by the relay can not be ignored in the downlink transmission. Secondly, the optimal relay location, which lies in the center of the contour, changes with the asymmetric traffic condition. The smaller the ratio of uplink traffic load to total traffic load is, the closer the optimal relay location is to MS.

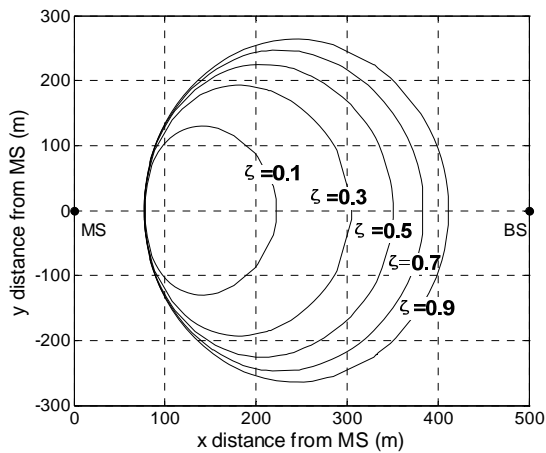

**Fig. 4** Energy-efficient cooperation regions for different asymmetric traffic factor $\zeta$, $D = 500$ m

## 5  Performance Evaluation

### 5.1  Diversity-multiplexing tradeoff of JUDRS

This subsection analyzes the diversity-multiplexing tradeoff (DMT) performance of the proposed relay selection scheme. The definition given in [16] is used. A channel is said to achieve multiplexing gain $r$ and diversity gain $d$ if there exists a sequence of codes $C(\rho)$ operating at SNR $\rho$ with rate $R(\rho)$ and resulting outage probability $P_{\text{out}}(\rho)$ such that:

$$\lim_{\rho \to \infty} \frac{R(\rho)}{\log_2(\rho)} = r, \qquad \lim_{\rho \to \infty} \frac{\log P_{\text{out}}(\rho)}{\log_2(\rho)} = -d \tag{14}$$

In the following developments, we say $f(\rho)$ is exponentially equal to $\rho^{\upsilon}$, denoted by $f(\rho) \doteq \rho^{\upsilon}$, if $\lim_{\rho \to \infty} \frac{\log_2 f(\rho)}{\log \rho} = \upsilon$.

The main DMT result for JUDRS scheme is given in the following theorem, where we denote $(\bullet)^+ = \max\{\bullet, 0\}$.

**Theorem 1 (DMT of JUDRS):** The JUDRS scheme achieves the following diversity-multiplexing tradeoff:

$$d_{\text{JUDRS}}(r) = (N+1)\left(1 - \frac{2N+1}{N+1}r\right)^+ \tag{15}$$

**Proof** Denote $\rho$ as the SNR without fading, which is defined as $\rho = \frac{P_{\max} - P_{\text{C}}}{N_0 B}$. The outage probability of the JUDRS scheme can be expressed as

$$P_{\text{out}} = \zeta \cdot P_{\text{out}}^{\text{UL}} + (1-\zeta) \cdot P_{\text{out}}^{\text{DL}} \tag{16}$$

where $P_{\text{out}}^{\text{UL}}$ and $P_{\text{out}}^{\text{DL}}$ are the outage probability of uplink and downlink, respectively. For the uplink transmission, during the broadcast phase, the mutual information across the MS-BS channel is

$$I_{\text{D}} = \log_2(1 + \rho h_{\text{D}}) \tag{17}$$

If a retransmission occurs, the combination of the two transmission forms an equivalent channel between the MS and BS, whose mutual information is:

$$I_{\text{2-hop}} = \log_2\left(1 + \rho\left(h_{\text{D}} + h_{2,i^*}\right)\right) \tag{18}$$

where $i^*$ denotes the index of the selected relay. Using the law of total probability, the outage probability of uplink transmission can be expressed as:

$$P_{\text{out}}^{\text{UL}} = \left(\sum_{t=1}^{N} \Pr\{I_{\text{2-hop}} < R \mid |\Sigma| = t, I_{\text{D}} < R\} \cdot \Pr\{|\Sigma| = t\} + \Pr\{|\Sigma| = 0 \mid I_{\text{D}} < R\}\right) \cdot \Pr\{I_{\text{D}} < R\} \tag{19}$$

where $R = r\log_2 \rho$ and $|\Sigma|$ denotes the cardinality of $\Sigma$. The probability that the direct link is in outage is given by

$$\Pr\{I_{\text{D}} < R\} = \Pr\left\{h_{\text{D}} < \frac{2^R - 1}{\rho}\right\} = \int_0^{\rho^{r-1} - \rho^{-1}} \frac{1}{KD^{-p}} \exp\left(\frac{x}{KD^{-p}}\right) dx \doteq \rho^{r-1} \tag{20}$$

And the condition probability that the 2-hop channel is in outage can be calculated as,

$$\Pr\{I_{\text{2-hop}} < R \mid I_{\text{D}} < R, |\Gamma| = t\} = \Pr\left\{\frac{1}{2}\log_2\left(1 + \rho(h_{\text{D}} + h_{2,i^*})\right) < R \mid \log_2(1 + \rho h_{\text{D}}) < R, i^* \in \Gamma\right\} = 0 \tag{21}$$

Here, Eq. (21) holds because

$$\frac{1}{2}\log_2\left(1 + \rho(h_{\text{D}} + h_{2,i^*})\right) \geq \frac{1}{2}\log_2\left(1 + \rho\left(h_{\text{D}} + \frac{2^{2R} - 1}{\rho}\left(1 - \frac{h_{\text{D}}}{h_{1,i^*}}\right)\right)\right) \geq R \tag{22}$$

The probability that no relay can help the transmission between BS and MS is given by

$$\Pr\{|\Sigma| = 0 \mid I_{\text{D}} < R\} = \sum_{\Gamma \subset \{1,2,\ldots,N\}} \prod_{i \in \Gamma} \Pr\left\{h_{2,i} < \frac{2^{2R} - 1}{\rho}\left(1 - \frac{h_{\text{D}}}{h_{1,i}}\right) \mid h_{\text{D}} < \frac{2^R - 1}{\rho}, h_{1,i} \geq \frac{2^R - 1}{\rho}\right\} \cdot \Pr\{\Gamma\} \tag{23}$$

The upper and lower bounds of the first term in Eq. (23) can be derived as

$$\Pr\left\{h_{2,i} < \frac{2^{2R} - 1}{\rho}\left(1 - \frac{h_{\text{D}}}{h_{1,i}}\right) \mid h_{\text{D}} < \frac{2^R - 1}{\rho}, h_{1,i} \geq \frac{2^R - 1}{\rho}\right\} \leq \Pr\left\{h_{2,i} < \frac{2^{2R} - 1}{\rho}\right\} \doteq \rho^{2r-1} \tag{24}$$

and

$$\Pr\left\{h_{2,i} < \frac{2^{2R} - 1}{\rho}\left(1 - \frac{h_{\text{D}}}{h_{1,i}}\right) \mid h_D < \frac{2^R - 1}{\rho}, h_{1,i} \geq \frac{2^R - 1}{\rho}\right\} > \Pr\left\{h_{2,i} < \frac{2^R(2^{2R} - 1)}{\rho(2^R + 1)}\right\} \doteq \rho^{2r-1} \tag{25}$$

The second term in Eq. (23) is given by

$$\Pr\{\Gamma\} = \prod_{i \in \Gamma} \Pr\left\{h_{1,i} \geq \frac{2^{2R} - 1}{\rho}\right\} \cdot \prod_{i \notin \Gamma} \Pr\left\{h_{1,i} < \frac{2^{2R} - 1}{\rho}\right\} \doteq \rho^{(2r-1)(N-t)} \tag{26}$$

where $t = |\Gamma|$. Thus, the outage probability of uplink transmission is derived as

$$P_{\text{out}}^{\text{UL}} \doteq \rho^{(2N-1)r - (N+1)} \tag{27}$$

The same methodology can be applied to the downlink. Therefore, the DMT of the JUDRS scheme is given by

$$d_{\text{JUDRS}}(r) = (N+1)\left(1 - \frac{2N+1}{N+1}r\right)^+ \tag{28}$$

∎

The DMT analysis corroborates the merits of the JUDRS scheme that it can achieve full spatial diversity in the number of the cooperating nodes, not just the number of decoding relays.

### 5.2 Energy-efficiency comparison

In this subsection, the energy consumption performance of the proposed JUDRS scheme is provided by simulation. For comparison, we take two well-known relay selection schemes, the best worst channel selection and the best harmonic mean selection [17] as the benchmark. In the best worse channel selection scheme, the relay whose worse channel, $\min\{h_{1,i}, h_{2,i}\}$, is the best is selected. In best harmonic mean selection, the relay selection function is chosen as the harmonic mean of the two channel's magnitudes: $\left(\left(h_{1,i}\right)^{-1}+\left(h_{2,i}\right)^{-1}\right)^{-1}$. The relay with the largest harmonic mean cooperates. We extend the best worse channel selection and the best harmonic mean selection to both uplink and downlink straightforward. The parameters are the same as in Table 1.

In Fig. 5(a) and Fig. 5(b), the total energy consumption per information bit of battery-powered terminals is plotted as a function of the number of relays for two scenarios: a direct link between MS and BS exists (see Fig. 5(a)}, MS-BS distance of 450m) and a direct link does not exists (see Fig. 5(b), MS-BS distance of 1200m, $h_D \ll \min\{h_{1,i}, h_{2,i}\}$). These figures indicate that the proposed JUDRS scheme can save the total energy efficiently compared with the best worse channel selection and the best harmonic selection in both cases with different traffic load conditions. It is also shown in the figures that the energy consumed for transmitting one bit decreases with the active relay number, which means that it is helpful to have a larger number of relays.

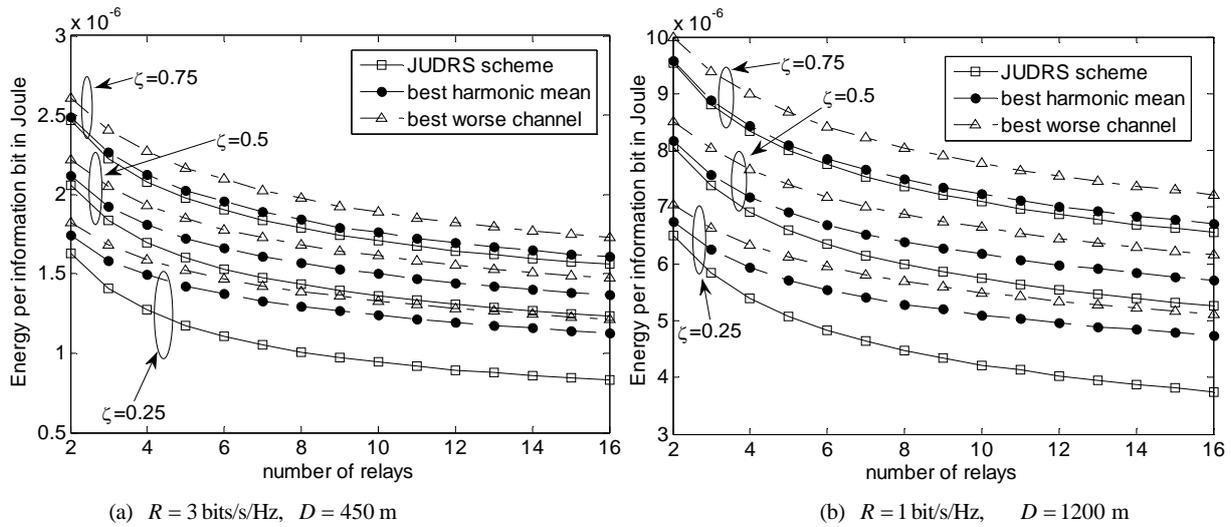

(a) $R = 3$ bits/s/Hz, $D = 450$ m   (b) $R = 1$ bit/s/Hz, $D = 1200$ m

**Fig. 5** Total energy consumption of MS and relay per information bit v.s. number of relays

Fig. 6 compares the energy consumption of MS and relay for transmitting one bit under different traffic conditions when 8 relays are active. It is shown that the energy saving achieved by the proposed scheme grows with the decrease of the asymmetric traffic factor $\zeta$. It gives the advantage of the JUDRS scheme that it can adaptively adjust the transmit energy to the minimum according to the traffic condition in the system. As the uplink traffic load is always less than that of the downlink, we can always expect a larger energy-saving using the JUDRS scheme. Fig. 6 also indicates the energy-saving achieved by the JUDRS scheme mainly comes from the energy saving of the relay, especially when the uplink traffic load is less than the downlink traffic load.

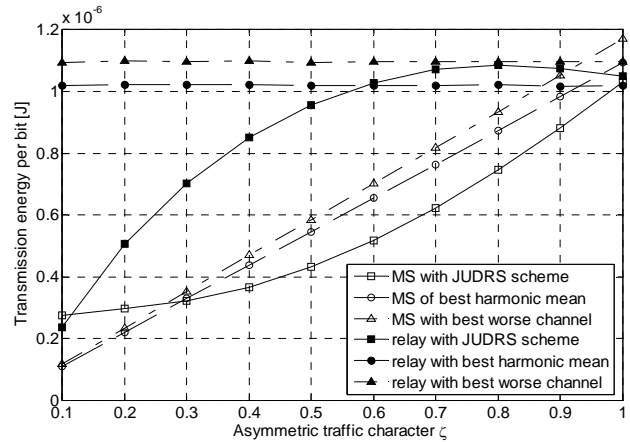

**Fig. 6** The energy consumption of MS and relay per bit for different traffic conditions, with 8 active relays and MS-BS distance of 450m, $R = 3$ bit/s/Hz.

## 6 Conclusions

In this paper, we propose an energy-efficient relaying scheme though selection of mobile relays in cooperative cellular systems with asymmetric traffic. Specifically, we generalize the proposed JUDRS scheme in two key aspects: (i) relay is selected jointly for the uplink and the downlink, so that the relay selection overhead is reduced dramatically, (ii) the total energy consumption per information bit of the battery-powered terminals is minimized while the channel quality and the traffic load condition of the uplink and downlink are considered. Afterwards, the energy-efficient cooperation regions are investigated, and the optimal relay location is found for cooperative cellular systems with asymmetric traffic. Information theoretic analysis of the diversity-multiplexing tradeoff (DMT) demonstrates that the proposed scheme achieves full spatial diversity in the quantity of cooperating terminals in this network. Finally, numerical results further confirms that the proposed algorithm can improve the energy-efficiency of the MS and the relay compared with the previous best worse channel selection and best harmonic mean selection algorithms.

**Acknowledgements**

The authors would like to thank Dr. Gang Wu and Dr. Haifeng Wang of Nokia for many helpful discussions. This paper is supported by the National Science Foundation of China (NSFC 60702051, NSFC-AF: 60910160), the Specialized Research Fund for the Doctoral Program of Higher Education (SRFDP：20070013028), and the Program for New Century Excellent Talents in University (NCET-08-0735). This paper is co-funded by Nokia on the Beyond 3G Research project.

## References

1. Sendonaris A, Erkip E, Aazhang B. User cooperation diversity - part I: system description. IEEE Transactions on Communications, 2003, 51(11): 1927-1938
2. Laneman J N, Tse D N C, Wornell G W. Cooperative diversity in wireless networks: efficient protocols and outage behavior. IEEE Transactions on Information Theory, 2004, 50(12): 3062-3080
3. Tannious R, Nosratinia A. Spectrally-efficient relay selection with limited feedback. IEEE Journal on Selected Areas in Communications, 2008, 26(8): 1419-1428
4. Lin Z, Erkip E, Stefanov A. Cooperative regions and partner choice in coded cooperative systems. IEEE Transactions on Communications, 2006, 54(7): 1323-1334
5. Nosratinia A, Hunter T E. Grouping and partner selection in cooperative wireless networks. IEEE Journal on Selected Areas in Communications, 2006, 54(4): 369-378
6. Madan R, Mehta N B, Molisch A F, et al. Energy-efficient cooperative relaying over fading channels with simple relay selection. IEEE Transaction on Wireless Communications, 2008, 7(8): 3013-3025
7. Bletsas A, Khisti A, Reed D R, et al. A simple cooperative diversity method based on network path selection. IEEE Journal on Selected Areas in Communications, 2006, 24(3): 659-672
8. Cui S, Goldsmith A J, Bahai A. Energy-constrained modulation optimization. IEEE Transactions on Wireless Communications, 2005, 4(5): 2349-2360
9. Cui S, Goldsmith A J, Bahai A. Energy-efficiency of MIMO and cooperative MIMO techniques in sensor networks. IEEE Journal on Selected Areas in Communications, 2004, 22(6):1089-1098
10. Goldsmith A J, Chua S G. Variable-rate variable-power MQAM for fading channels. IEEE Transaction on Communications, 1997, 45(10), pp: 1218-1230
11. Proakis J G. Digital communications (fourth edition), USA: McGraw-Hill Science Engineering, 2000: 511-526
12. Vandendorpe L, Louveaux J, Oguz O, et al. Power allocation for OFDM Transmission with DF Relaying. Proceedings of IEEE International Conference on Communications (ICC 2008), May 19-23, 2008, Beijing, China. 2008: 3795--3800
13. Miao G, Himayat N, Li Y et al. Energy-efficient design in wireless OFDMA. Proceedings of IEEE International Conference on Communications (ICC 2008), May 19-23, 2008, Beijing, China. 2008: 3307--3312
14. 3GPP TR 36.942 V8.2.0 (2009-5), Evolved Universal Terrestrial Radio Access (E-UTRA); Radio Frequency (RF) system scenarios, Release 8
15. Cho W, Oh W S, Kwak D Y. Effect of relay locations in cooperative networks. Proceedings of 1st International Conference on Wireless Communication, Vehicular Technology, Information Theory and Aerospace & Electronic Systems Technology (Wireless VITAE 2009), May 17-20, 2009, Aalborg, Denmark. 2009: 737--741.
16. Zheng L, Tse D N C. Diversity and multiplexing: a fundamental tradeoff in multiple-antenna channels. IEEE Transactions on Information Theory, 2003, 49(5):1073-1096
17. Jing Y, Jafarkhani H. Single and multiple relay selection schemes and their achievable diversity orders. IEEE Transactions on Wireless Communications, 2009, 8(3): 1414-1423